\documentclass[runningheads]{llncs}
\usepackage[T1]{fontenc}
\usepackage{graphicx}
\usepackage{xcolor}
\usepackage{bm}
\usepackage{hyperref}

\begin{document}
%
%\title{Improving Liver Segmentation through Multimodal Learning using CBCT and CT Scans}
%\title{CT And CBCT Based Multimodal Learning For Improving Semantic Segmentation}
%\title{Multimodal Learning With Variably Aligned Data To Improve Segmentation With Intraoperative CBCT \& Preoperative CT}
\title{Multimodal Learning With Intraoperative CBCT \& Variably Aligned Preoperative CT Data To Improve Segmentation}
\titlerunning{Multimodal Learning for Liver Segmentation}

\author{
Maximilian E. Tschuchnig\inst{1}\inst{2}\orcidID{0000-0002-1441-4752} \and 
Philipp Steininger\inst{3} \and 
Michael Gadermayr\inst{1}\orcidID{0000-0003-1450-9222} 
}
%\author{
%blind blind\inst{1}\inst{2}\orcidID{blind-blind-blind-blind} \and 
%blind blind\inst{3} \and 
%blind blind\inst{1}\orcidID{blind-blind-blind-blind} 
%}

\authorrunning{M. Tschuchnig et al.}
%\authorrunning{blind}

\institute{
Salzburg University of Applied Sciences 
\email{\{firstname,lastname\}@fh-salzburg.ac.at} \and
University of Salzburg \and
MedPhoton GmbH\\
}

%\institute{
%blind 
%\email{\{firstname,lastname\}@blind.com} \and
%blind \and
%blind\\
%}

%
\maketitle              % typeset the header of the contribution
\begin{abstract}
Cone-beam computed tomography (CBCT), is an important tool facilitating computer aided interventions, despite often suffering from artifacts that pose challenges for accurate interpretation. While the degraded image quality can affect downstream segmentation, the availability of high quality, preoperative scans represents potential for improvements. Here we consider a setting where preoperative CT and intraoperative CBCT scans are available, however, the alignment (registration) between the scans is imperfect. 
We propose a multimodal learning method that fuses roughly aligned CBCT and CT scans and investigate the effect of CBCT quality and misalignment on the final segmentation performance. For that purpose, we make use of a synthetically generated data set containing real CT and synthetic CBCT volumes. As an application scenario, we focus on liver and liver tumor segmentation. 
We show that the fusion of preoperative CT and simulated, intraoperative CBCT mostly improves segmentation performance (compared to using intraoperative CBCT only) and that even clearly misaligned preoperative data has the potential to improve segmentation performance. 
\end{abstract}
\section{Introduction}
% Motivation - intraoperative CBCT. Should be quick and as good as possible, does not have to be perfect but quick. Help through CT

To establish computer-assisted interventions, precise and reliable imaging, especially intraoperative imaging, is crucial. Mobile robotic medical imaging systems, like cone-beam computed tomography (CBCT)~\cite{rafferty2006intraoperative}, enable intraoperative medical imaging with real time capabilities.
% Mobile CBCT has gained popularity in intraoperative imaging due to its portability and real-time capabilities, despite its tendency to exhibit lower visual quality and artifacts compared to CT. 
CBCT is an imaging method that utilizes a cone-shaped X-ray beam and a flat-panel detector to capture detailed, three-dimensional images of a patient's anatomy using a mobile system~\cite{jaffray2002flat}.
However, this type of intraoperative imaging often comes with the disadvantage suffering from more artifacts than preoperative CT imaging, affecting the performance of downstream tasks like segmentation~\cite{wei2024reduction}.

% Introduction Multimodal Learning
While this degraded image quality can affect downstream segmentation, the availability of high quality preoperative scans represents a high potential for improvements based on the idea of multimodal learning~\cite{zhang2021deep,podobnik2023multimodal,zhang2021modality}.
Multimodal learning is an approach that involves fusing images from multiple domains to improve machine learning models for a downstream task like segmentation. In medical imaging, a common approach is to enrich computed tomograpy (CT) data, focusing on bone structures, with magnetic resonance imaging data for soft tissue analysis~\cite{zhang2021deep,podobnik2023multimodal}. Multimodal learning is typically separated into three fusion strategies~\cite{zhang2021deep,zhang2021modality}: early, late and hybrid. 
% early
The most common multimodal fusion, early-fusion, combines images of different modalities before being processed by a downstream model. Typically, the two domains are fused along a dimension additional to the spacial volume dimensions and processed jointly~\cite{zhang2021deep,ren2021comparing}. Another form of early fusion processes the volumes of different modalities in separate feature extraction stages, finally fusing the extracted features. % In the case of segmentation using unet, this corresponds to fusing the output of multiple encoders in the latent space and a shared decoder. 
% late
Late-fusion is performed before the final layer of the downstream task. In a segmentation case, late-fusion merges the features extracted from multiple independent encoder-decoder networks before the final layer (facilitating segmentation). Hybrid-fusion, combines aspects of both early and late fusion for enhanced performance. 

% Combined with registration
Typically, multimodal learning assumes that the fused images are aligned, utilizing affine or even elastic registration. However, registration of 3D samples, if performed accurately on a high resolution, is computationally expensive, especially if non-linear deformations need to be compensated.
Deep learning-based approaches~\cite{balakrishnan2019voxelmorph,chen2022transmorph} have shown promising performance while also reducing computational complexity, compared to classical, optimization based approaches.
Podobnik et al.~\cite{podobnik2023multimodal} integrated registration within their hybrid multimodal segmentation approach. Integrated registration was achieved by merging features from different modality branches using the affine Spatial Transformer Networks (STN) localization net~\cite{jaderberg2015spatial}, along with a grid generator and sampler.

% Hypothesis
Here we consider the setting where preoperative CT scans and intraoperative CBCT scans are available, however, the alignment (registration) between the scans is imperfect. 
We hypothesised that, by adding high quality, preoperative CT to intraoperative CBCT scans, segmentation performance will increase. We assume that, the more accurate the alignment between the modalities is, the more pronounced the performance increase will be. 

\textbf{Contribution: } 
We propose a method that combines roughly aligned CBCT and CT scans (early-fusion) and investigate the effect of CBCT quality and misalignment (based on affine and elastic transformations) on segmentation performance.

In detail, we synthesised a collection of synthetic CBCT data, focusing on the segmentation of liver and liver tumors based on the LiTS CT dataset~\cite{bilic2023liver}. Beyond varying the amount of digitally reconstructed radiographs (DRR) used for CBCT synthesis, resulting in $5$ different imaging qualities, we generated $9$ variably misaligned versions based on linear and non-linear models and $1$ baseline model resulting in overall $50$ sub data sets. We evaluated all $45$ combinations based on a unet architecture and compared the performances to the $5$ unimodal baseline settings.

\section{Methodology}
% First, give overview
We propose to fuse intraoperative CBCT with roughly aligned, high quality, preoperative CT to investigate if this kind of additional information improves model training for computer aided intervention systems.
% Maybe add something similar to some multimodal related work here
To accomplish this study we apply multimodal learning in medical imaging for the downstream task of liver and liver tumor segmentation using CT and CBCT data based on the LiTS dataset~\cite{bilic2023liver}\footnote{For the CBCT version of LiTS see the Kaggle dataset \textit{CBCT Liver and Liver Tumor Segmentation Train Data}}. Similar to Ren et al.~\cite{ren2021comparing} we used early-fusion for the multimodal setup. Furthermore, we investigated two parameters, $\alpha_{np}$ as the number of DRRs used to simulate the current CBCT (representing undersampling), and $\alpha_a$ as the current alignment factor between the preoperative CT and the intraoperative CBCT. 
% Mean Dice of the CBCT segmentation targets of either liver or liver tumor over $4$ repetitions of each experiment was used as the evaluation metric.
As a baseline, segmentation was performed using only the intraoperative CBCT scans, which was evaluated for all settings of $\alpha_{np}$. Applied to the task of liver and liver segmentation, this setup results in $100$ different experiments. Misalignment was performed using affine transformations during training and validation to decrease computational complexity in the dataloader. During validation, which is only performed once for each sample, affine misalignment was followed by elastic misalignment to evaluate the models ability to generalize to non-linear deformations.

\begin{figure}[ht]
    \centering
    \includegraphics[width=1\textwidth]{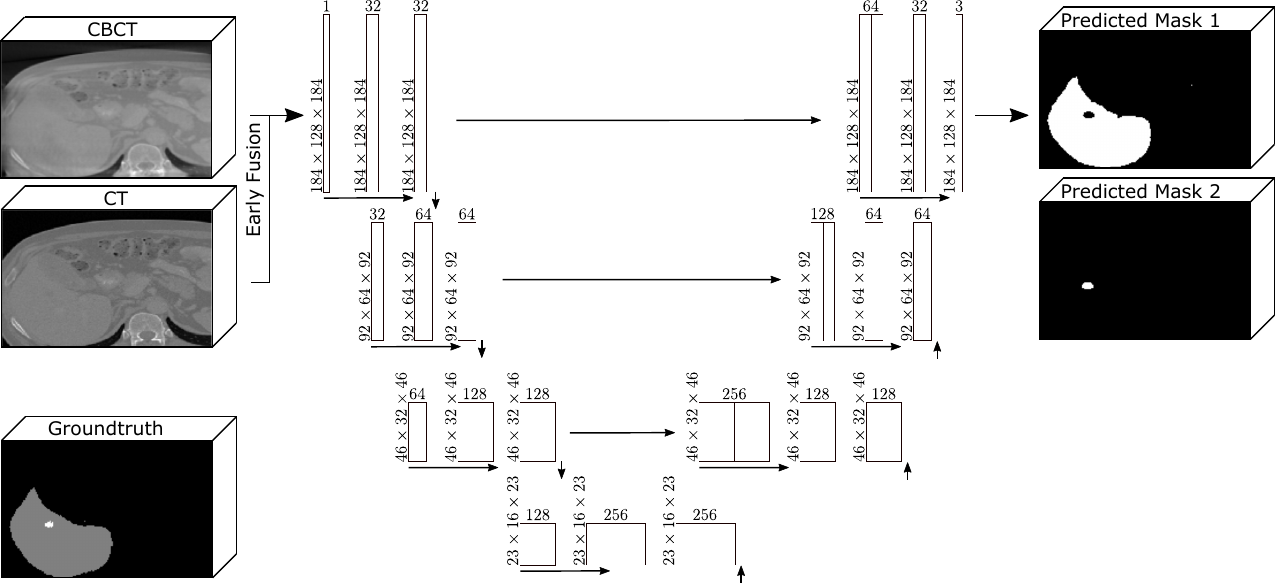}
    \caption{Multimodal model configuration. After fusing the intraoperative CBCT and preoperative CT (early fusion), the data was processed by the shown unet, segmenting liver and liver tumors.}
    \label{fig:model}
\end{figure}

% Show model in more detail but no experimental details
We used a holistic, 3D unet introduced by Çiçek et al.~\cite{cciccek20163d} as the basis of our segmentation model and our baseline for all investigated settings. The 3D unet was updated to process the multimodal data by adding a paired and variably misaligned, preoperative CT as a second channel, resulting in a 4d data structure.
% Give info about unet and its specific learning parameters
The 3D unet used for segmentation, shown in Fig.~\ref{fig:model}, consisted of an encoder with 3 double convolution layers and $3 \times 3 \times 3$ convolutional kernels, connected by 3D max pooling. The latent space was constructed using one double convolution block followed by the unet decoder, mirroring the encoder. As is typical for unet, each double convolutional output in the encoder was also connected to the decoder double convolutional block of the same order. Additionally, one 3D convolutional layer was added to the decoder with a filter size of $1 \times 1 \times 1$ and the number of filters set to the number of segmentation classes (in the case of liver and liver tumor segmentation this value was set to $2$). The number of feature maps were set to $\{32, 64, 128, 256\}$ as shown in Fig.~\ref{fig:model}. Batch norm was applied after each layer in the double convolutional blocks. The model was trained utilizing a sum of binary cross-entropy and Dice similarity. For our baseline, a unimodal unet was used, with only the CBCT as input. Our multimodal approach added the high quality, misaligned and preoperative CT as a second channel to the CBCT.

\subsection{Experimental Details}
% Give infos about our deep learning server and grafics card a6000
All models were trained on an Ubuntu server using NVIDIA RTX A6000 graphics cards. Due to the large size of the data and memory restrictions (48 GB VRAM), the volumes were downscaled (isotropic) by the factor of two. 
% Show evaluation criteria
To binarize the masks, a threshold of $0.5$ was applied to each channel of the unet output. As an additional baseline, CBCT volumes with perfectly aligned CT volumes, $\alpha_a=0$, was also investigated. All experiments were trained and evaluated $4$ times to facilitate stable results with the same random splits as well as the same random CT misaligned for comparable result. Adam was used as an optimizer with a learning rate of $0.005$. The source code for the experiments can be found on \url{https://github.com/blind/blind}.
% Show datasplits
Since annotations were not available for the LiTS test dataset, the test dataset was disregarded for this publication and the (processed) LiTS training set was separated into training-validation-testing data~\cite{araujo2022liver,han2022effective}. The separation was performed using the ratios $0.7$ (training), $0.2$ (validation), $0.1$ (testing).

\subsection{Dataset Generation}
For evaluation of the multimodal learning approach, paired CT and CBCT volumes with variable degrees of misalignment had to be generated.

\begin{figure}[htb]
\includegraphics[width=\textwidth]{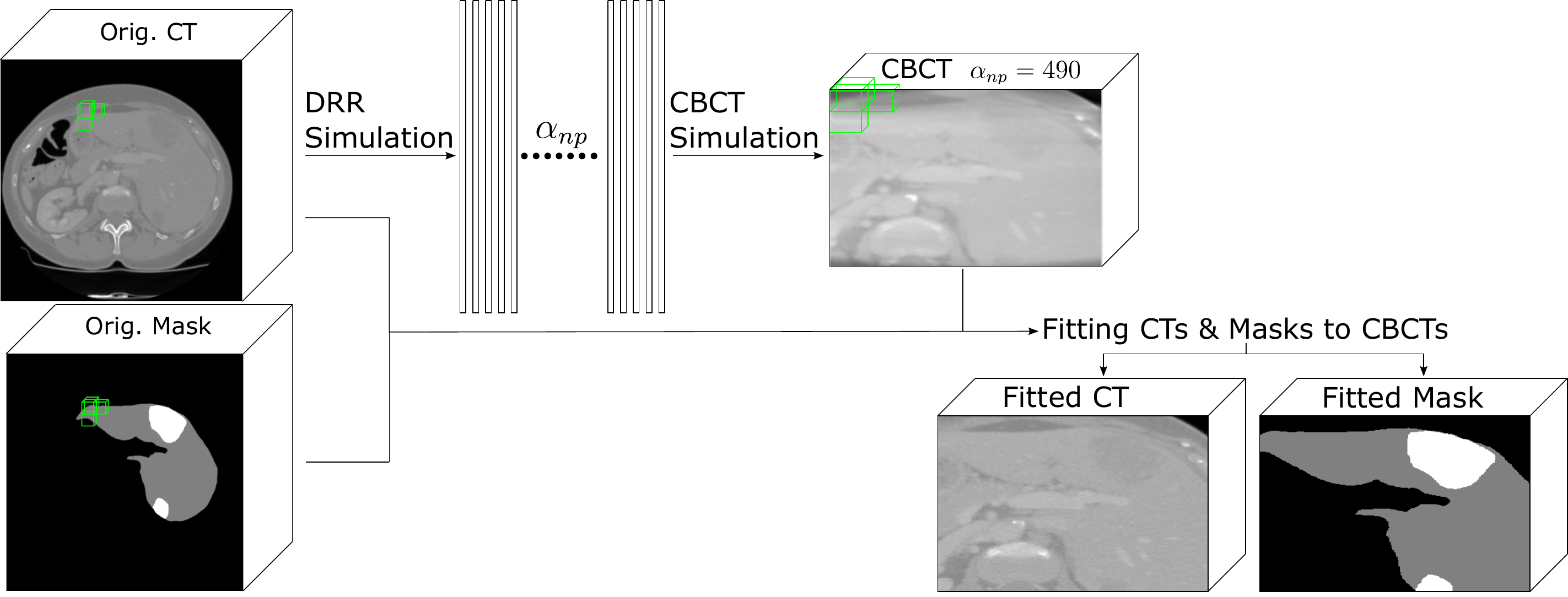}
\caption{
Data generation process: after centering the original CT volumes around the liver (using the liver segmentations), $\alpha_{np}$ projections were simulated. Finally, CBCT were simulated and aligned with the original CT and masks, in order to fit them to the CBCT field-of-view.} \label{fig:datapreprocessing}
\end{figure}

% Shortly explain the dataset and how the CBCT came to be (use BVM as source)
To generate CBCT/CT pairs we simulated DRRs from the CT volumes, with variable undersampling (with $\alpha_{np} \in \{32, 64, 128, 256, 490\}$ representing the predefined number of DRRs). We then used these DRRs to simulate CBCT with varying visual quality. Higher $\alpha_{np}$ corresponded to better image quality, with $\alpha_{np}=490$ serving as the CBCT quality similar to preoperative CT and $\alpha_{np}=32$ as the lowest visual quality with significant artifacts~\cite{tschuchnig2024multi}.
Fig.~\ref{fig:datapreprocessing} shows these steps to convert the CT LiTS dataset into synthetic CBCT scans. % It also displays how corresponding CT and masks were fitted to the shapes of the synthesised CBCT. 

The LiTS dataset was chosen to perform the experiments~\cite{bilic2023liver}. LiTS consists of $131$ abdominal CT scans in the training set and $70$ test volumes. The $131$ training volumes include segmentations of 1) the liver and 2) liver tumors. The dataset contains data from $7$ different institutions with a diverse set of liver tumor diseases. 
%These diseases include primary and secondary liver tumors with varying lesion-to-background ratios. There is also a mix of pre- and post-therapy CT scans.
The CT scans were acquired using different CT scanners and acquisition protocols. For further information about the dataset we refer to Bilic et al.~\cite{bilic2023liver}\footnote{To download the LiTS dataset follow the link: \url{https://competitions.codalab.org/competitions/17094}}. 

% Explain the basic data unalignment process
Affine misaligned was performed using random (non-isotropic) scaling, with the scaling parameter sampled from $\mathcal{U}(1 - 0.5 \cdot \alpha_a,1 + 0.5 \cdot \alpha_a)$, rotation, parameter sampled from $\mathcal{U}(-22.5 \cdot \alpha_a, 22.5 \cdot \alpha_a)$, and translation, with the parameter sampled from $\mathcal{U}(0, 0.5 \cdot \alpha_a)$ with tri-linear interpolation. Elastic misalignment was applied with a maximum displacement sampled from $\mathcal{U}(0, 20 \cdot \alpha_a)$ with $7$ control points and no locked borders.
%For other affine and elastic deformation parameters the default configuration was chosen. 
To reduce the number of parameters controlling misalignment parameters to one, the alignment factor $\alpha_a \in \{0, 0.125, 0.25, 0.5, 1\}$ was introduced, controlling the misalignmnet between pairs of scans. Augmentation was performed using TorchIO RandomAffine and RandomElasticDeformation.

\begin{figure}
\includegraphics[width=\textwidth]{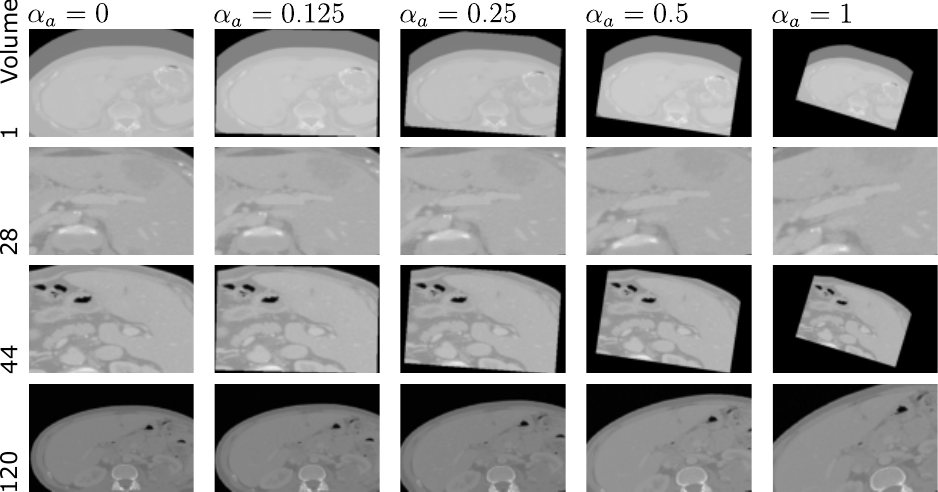}
\caption{Results of the random affine augmentation with differing augmentation factor $\alpha_a \in \{0, 0.125, 0.25, 0.5, 1\}$ of $4$ different volumes.} \label{fig:augresultingdata}
\end{figure}

\begin{figure}
\includegraphics[width=\textwidth]{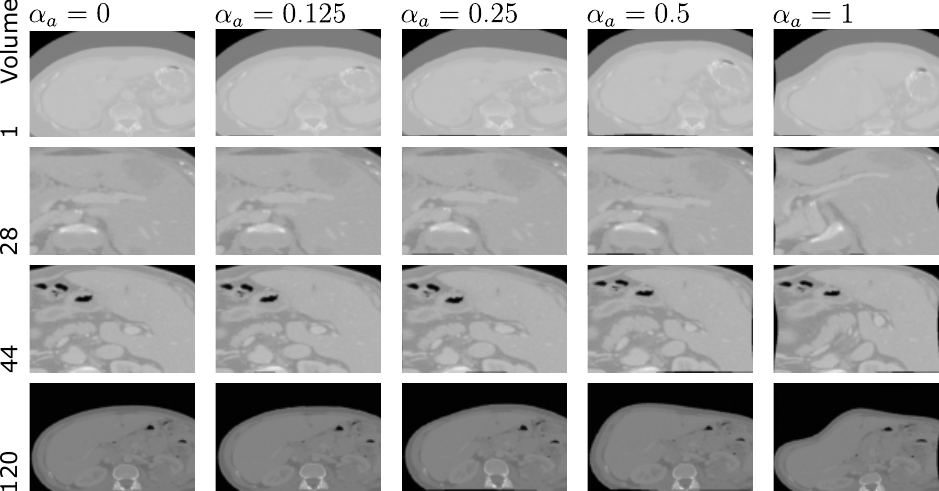}
\caption{Results of the random elastic augmentation with differing augmentation factor $\alpha_a \in \{0, 0.125, 0.25, 0.5, 1\}$ of $4$ different volumes.} \label{fig:augresultingdataelastic}
\end{figure}

Fig~\ref{fig:augresultingdata} shows examples of affine misalignment with the $\alpha_a \in \{0, 0.125, 0.25, 0.5, 1\}$ and $4$ different volumes. Fig~\ref{fig:augresultingdataelastic} shows the same volumes, with applied random elastic deformation. Here we display the elastic transformation in isolation from affine transformations for improved visibility.

% Show images of CT, CBCT and deformed data

% Show images of elastic deformed data

\section{Results}
Experimental results are shown in Table~\ref{tab:resultsAll}, denoted as mean Dice values. If the test data was transformed using the affine transformations, rows are marked as \textit{affine-s}. Affine, followed by the elastic transformed settings are marked as \textit{elastic-s}. Two baselines are also shown: unimodal CBCT with no preoperative CT (\textit{base CBCT}) and CBCT with perfectly aligned CT (\textit{no misalignment}). Blue values show improvements compared to the baseline (\textit{base CBCT}), while red values show decreased scores. If the values are bold there is at least a $5\%$ increase/decrease. The $+,-$ corresponds to improvements/decreases to the previous, less accurately aligned experiments. If $+,-$ are colored, the increase/decrease compared to the previous less accurately aligned example is larger than $5\%$. The different undersampled $\alpha_{np}$ are displayed horizontally and $\alpha_a$ vertically for both liver and liver tumor segmentation.
In the affine setting, the scores increased compared to the baseline in all liver segmentation and in $11$ out of $20$ liver tumor segmentation cases. The biggest improvement for liver segmentation was achieved with the parameters $\alpha_{np}=32$ and $\alpha_a=0$ with an improvement from $0.784$ to $0.932$ resulting in an increase of $0.148$. For liver tumor segmentation, the biggest improvement was also achieved with the parameters $\alpha_{np}=32$ and $\alpha_a=0$ from $0.029$ to $0.297$ resulting in an increase of $0.268$. There was also an observable trend of increasing Dice scores from $\alpha_a \in \{0.5, 0.25, 0.125, 0\}$. This trend was stronger, the lower $\alpha_{np}$ was.
Adding elastic transformation on top of affine transformations lead to slightly lower scores in the cases of $\alpha_{a} \in {0.5, 0.25}$ with $\alpha_{a=0.25}$ suffering most from adding elastic transformations with an average Dice decrease of $0.023$ for liver and $0.032$ for liver tumor segmentation. The elastic transformation effects regarding $\alpha_{np}$ show a decrease of average Dice proportional to image quality. The most significant decrease in Dice, attributable to elastic transformation was reached with $\alpha_{np}=32$ with $0.016$ for liver and $0.021$ for liver tumor segmentation.

% Show results as table
\begin{table}[htb] % only baseline + affine + elastic
\caption{Experimental results (affine and elastic misalignment) showing the mean Dice values. Blue values denote improvements compared to the baseline (\textit{base CBCT}), while red values show decreased scores. If the values are in bold there is at least a $5\%$ increase/decrease. The $+,-$ corresponds to increases/decreases compared to the previous, less accurately aligned setting. If $+,-$ are colored, this increase/decrease is larger than $5\%$.} \label{tab:resultsAll}
\resizebox{\textwidth}{!}{\begin{tabular}{l|llllllllll|llllllllll}
& 
\multicolumn{10}{c}{\textbf{Liver Segmentation}} & 
\multicolumn{10}{c}{\textbf{Liver Tumor Segmentation}} \\
&
\textbf{490} & & 
\textbf{256} & &
\textbf{128} & &
\textbf{64} & &
\textbf{32} & &
\textbf{490} & &
\textbf{256} & &
\textbf{128} & &
\textbf{64} & &
\textbf{32} & \\ \hline

\textbf{base CBCT} & 
0.884 & 
& 0.884 &
& 0.859 &
& 0.817 &
& 0.784 &
& 0.165 &
& 0.162 &
& 0.093 &
& 0.061 & 
& 0.029 & \\ 

\textbf{no misalignment} & 
\color{blue}0.933\color{black} & $\bm{+}$ & 
\color{blue}0.931\color{black} & $\bm{+}$ & 
\textbf{\color{blue}0.931} & $\bm{+}$ & 
\textbf{\color{blue}0.933} & $\bm{+}$ & 
\textbf{\color{blue}0.932} & $\bm{+}$ & 
\textbf{\color{blue}0.330} & $\color{blue} \bm{+}$ & 
\textbf{\color{blue}0.322} & $\color{blue} \bm{+}$ & 
\textbf{\color{blue}0.298} & $\color{blue} \bm{+}$ & 
\textbf{\color{blue}0.325} & $\color{blue} \bm{+}$ & 
\textbf{\color{blue}0.297} & $\color{blue} \bm{+}$ \\ \hline

\textbf{affine-s1} & 
\color{blue}0.906\color{black} & $\bm{+}$ & 
\color{blue}0.897\color{black} & $\bm{+}$ & 
\color{blue}0.879\color{black} & $\bm{+}$ & 
\color{blue}0.857\color{black} & $\bm{+}$ & 
\color{blue}0.806\color{black} & $\bm{+}$ &
\color{red}0.154\color{black} & $\bm{-}$ & 
\color{red}0.124\color{black} & $\bm{-}$ & 
\color{red}0.057\color{black} & $\bm{-}$ & 
\color{red}0.023\color{black} & $\bm{-}$ & 
\color{blue}0.051\color{black} & $\bm{+}$ \\

\textbf{affine-s0.5} & 
\color{blue}0.895\color{black} & $\bm{-}$ & 
\color{blue}0.891\color{black} & $\bm{-}$ & 
\color{blue}0.875\color{black} & $\bm{-}$ & 
\color{blue}0.863\color{black} & $\bm{+}$ & 
\textbf{\color{blue}0.851} & $\bm{+}$ & 
\textbf{\color{red}0.096} & $\color{red} \bm{-}$ & 
\textbf{\color{red}0.098} & $\bm{-}$ & 
\color{red}0.077\color{black} & $\bm{+}$ & 
\color{red}0.056\color{black} & $\bm{+}$ & 
\textbf{\color{blue}0.107} & $\color{blue} \bm{+}$ \\

\textbf{affine-s0.25} & 
\color{blue}0.904\color{black} & $\bm{+}$ & 
\color{blue}0.906\color{black} & $\bm{+}$ & 
\color{blue}0.893\color{black}& $\bm{+}$ & 
\textbf{\color{blue}0.883} & $\bm{+}$ & 
\textbf{\color{blue}0.883} & $\bm{+}$ &
\color{blue}0.181\color{black} & $\color{blue} \bm{+}$ & 
\color{red}0.160\color{black} & $\color{blue} \bm{+}$ & 
\textbf{\color{blue}0.163} & $\color{blue} \bm{+}$ & 
\textbf{\color{blue}0.169} & $\color{blue} \bm{+}$ & 
\textbf{\color{blue}0.171} & $\color{blue} \bm{+}$ \\

\textbf{affine-s0.125} & 
\color{blue}0.908\color{black} & $\bm{+}$ & 
\color{blue}0.900\color{black} & $\bm{-}$ & 
\color{blue}0.889\color{black} & $\bm{-}$ & 
\textbf{\color{blue}0.884} & $\bm{+}$ &
\textbf{\color{blue}0.887} & $\bm{+}$ & 
\color{blue}0.187\color{black} & $\bm{+}$ & 
\color{blue}0.166\color{black} & $\bm{+}$ & 
\textbf{\color{blue}0.166} & $\bm{+}$ & 
\textbf{\color{blue}0.152} & $\bm{-}$ & 
\textbf{\color{blue}0.185} & $\bm{+}$ \\ \hline

\textbf{elastic-s1} & 
\color{blue}0.907\color{black} & $\bm{+}$ & 
\color{blue}0.897\color{black} & $\bm{+}$ & 
\color{blue}0.880\color{black} & $\bm{+}$ & 
\color{blue}0.857\color{black} & $\bm{+}$ & 
\color{blue}0.081\color{black} & $\bm{+}$ &
\color{red}0.153\color{black} & $\bm{-}$ & 
\color{red}0.122\color{black} & $\bm{-}$ & 
\color{red}0.057\color{black} & $\bm{-}$ & 
\color{red}0.022\color{black} & $\bm{-}$ & 
\color{blue}0.049\color{black} & $\bm{+}$ \\

\textbf{elastic-s0.5} & 
\color{blue}0.895\color{black} & $\bm{-}$ & 
\color{blue}0.891\color{black} & $\bm{-}$ & 
\color{blue}0.870\color{black} & $\bm{-}$ & 
\color{blue}0.845\color{black} & $\bm{-}$ & 
\color{blue}0.800 & $\bm{+}$ & 
\textbf{\color{red}0.098} & $\color{red} \bm{-}$ & 
\textbf{\color{red}0.103} & $\bm{-}$ & 
\color{red}0.085\color{black} & $\bm{+}$ & 
\color{red}0.029\color{black} & $\bm{+}$ & 
\color{blue}0.049 & $\bm{+}$ \\

\textbf{elastic-s0.25} & 
\color{blue}0.893\color{black} & $\bm{-}$ & 
\color{blue}0.898\color{black} & $\bm{+}$ & 
\color{blue}0.869\color{black}& $\bm{-}$ & 
\color{blue}0.853 & $\bm{+}$ & 
\textbf{\color{blue}0.846} & $\bm{+}$ &
\color{red}0.154\color{black} & $\bm{+}$ & 
\color{red}0.142\color{black} & $\color{blue} \bm{+}$ & 
\color{blue}0.136 & $\color{blue} \bm{+}$ & 
\textbf{\color{blue}0.129} & $\color{blue} \bm{+}$ & 
\textbf{\color{blue}0.122} & $\color{blue} \bm{+}$ \\

\textbf{elastic-s0.125} & 
\color{blue}0.910\color{black} & $\bm{+}$ & 
\color{blue}0.901\color{black} & $\bm{+}$ & 
\color{blue}0.891\color{black} & $\bm{+}$ & 
\textbf{\color{blue}0.885} & $\bm{+}$ & 
\textbf{\color{blue}0.887} & $\bm{+}$ & 
\color{blue}0.186\color{black} & $\bm{+}$ & 
\color{blue}0.168\color{black} & $\bm{+}$ & 
\textbf{\color{blue}0.163} & $\bm{+}$ & 
\textbf{\color{blue}0.151} & $\bm{+}$ & 
\textbf{\color{blue}0.168} & $\bm{+}$ \\
\end{tabular}}
\end{table}

\section{Discussion}
% We hypothesise that by adding high quality, preoperative CT to our intraoperative CBCT, dice scores will increase the better the alignment is and that this effect is bigger, the worse the CBCT image quality is.
% Discuss the general results
The results confirmed the hypothesis that enriching intraoperative CBCT with roughly aligned preoperative CT can improve downstream tasks like segmentation. Most multimodal setups improved downstream performance with the only outliers observable in liver tumor segmentation with $\alpha_a \in \{0.5, 1\}$ and $\alpha_{np} \in \{490, 256, 128, 64\}$ and $(\alpha_a, \alpha_{np}) = (0.25, 256)$ in the affine misaligned cases. If data was additionally misaligned through elastic transformations, the settings $(\alpha_a, \alpha_{np}) = (0.25, 490)$ also lead to a decrease in average Dice.

% Discuss that in the worst image qualities, the effect is best, which is very useful for CBCT.
Several trends are observable. For one, the worse the CBCT quality, the more can be gained by adding high quality CT, leading to an increase of Dice from $0.784 \rightarrow 0.932$ for liver and $0.029 \rightarrow 0.297$ for liver tumor segmentation. There is a particularly pronounced effect in the case of adding preoperative information to low-quality interoperative CBCT. This effect is especially pronounced in the case of liver segmentation, where baseline scores can be reached using $\alpha_{np}=32$.

We also observe a trend regarding alignment of preoperative CT and intraoperative CBCT. At first, increased alignment lead to mostly worse average Dice for $\alpha_a \in \{1, 0.5\}$ followed by a notable increase in average Dice for $\alpha_a \in \{0.25, 0.125, 0\}$. This trend was only observed for liver tumor segmentation. We hypothesise that this is the case due to heavily misaligned data interfering with the training process. This is especially the case for complex structures like tumors. Here, the model can easily learn to ignore the higher quality but heavily misaligned CT but overfit to better aligned CT data. However, further experiments to determine the reason for this should be performed. For now our findings suggest that if there is significant misalignment and the target is difficult, such as liver tumor segmentation, preregistration is crucial for successful multimodal learning. 

The addition of elastic misalignment lead to a proportional decrease in downstream segmentation performance depending on the undersampling $\alpha_{np}$. Therefore, the same elastic misalignment had more pronounced degrading effects on segmentation performance the worse the image quality got, hinting at a possible implicit registration effect that diminishes with image quality. Investigating the factor $\alpha_a$, elastic misalignment had no effect in the cases of $\alpha_a \in \{1, 0.125\}$ with the average change $< 0.005$. However, in the cases of $\alpha_a \in \{0.5, 0.25\}$ the effect of adding elastic misalignment was on average a decrease of $0.021$. These insights hint at the deformation of $\alpha_a = 1$ already being almost unfeasible to handle in the affine case, leaving the addition of elastic transform in these cases without effect. Since we also observed almost no effect in the cases of $\alpha_a = 0.125$ we suspected that in these cases the elastic transform was negligible, in comparison to the affine misalignment. Interestingly, these performance gains show that the unet used for segmentation implicitly learned limited registration.

% pretty much a conclusion
% Similar to Zhang et al.~\cite{zhang2021modality} and Podobnik et al.~\cite{podobnik2023multimodal} we see huge potential in multimodal learning to improve intraoperative imaging with preoperative (imaging) information. Using multimodal learning, the negative effect of low intraoperative CBCT image quality could be greatly reduced for the downstream task of segmentation.  Assuming affine preregistration, this is applicable to a multitude of intraoperative imaging tasks e.g. to segment organs, lesions or metal, in order to remove it in further reconstruction steps, reducing imaging artifacts.

Although we only evaluated our results on unet, this form of multimodal learning is theoretically applicable to segmentation models based on similar building blocks, most importantly convolutional filters and pooling. Further experiments are needed to investigate this assumption for other architectures, such as the Segment Anything Model~\cite{kirillov2023segment} or UNETR~\cite{hatamizadeh2022unetr}. 

\section{Conclusion}
This study highlights the effectiveness of multimodal learning for downstream tasks, combining roughly aligned intraoperative CBCT with high quality, preoperative CT data.
% We showed that, using multimodal learning, performance can be improved without explicitly incorporating the fact that the alignment is imperfect.
It further investigates how the different factors of volume quality and volume alignment influence the performance of a specific multimodal learning based model. Using the multimodal learning setup, improvements in segmentation accuracy, especially when CBCT volume quality was suboptimal, were reported suggesting that high-quality preoperative CT data can compensate for intraoperative CBCT limitations, as long as the data is roughly aligned. Therefore, using this approach, clinicians could be supplied with more reliable information for surgical decision-making, particularly in near real-time settings of computer-assisted intervention. Assuming there is an efficient way for rough preregistration, this underscores the practical applicability of the approach. We further showed that a simple 3D unet was able to learn limited, implicit registration.

Since evaluation was based on synthetic data with relatively simple misalignment and without preregistration useful next steps would be to evaluate using real, paired preoperative and intraoperative data as well as incorporating (pre)registration into the multimodal model. Furthermore, continued exploration into multimodal approaches, including late and hybrid fusion, are promising future research directions.  
%In summary, multimodal learning offers promising potential to enhance intraoperative imaging, empowering clinicians with better tools for precise diagnosis, treatment planning, and surgical intervention.

\begin{credits}
\subsubsection{\ackname} This project was partly funded by the Austrian Research Promotion Agency (FFG) under the bridge project "CIRCUIT: Towards Comprehensive CBCT Imaging Pipelines for Real-time Acquisition, Analysis, Interaction and Visualization" (CIRCUIT), no. 41545455 and by the county of Salzburg under the project AIBIA.
% \subsubsection{\ackname} Blinded Funding. 

\subsubsection{\discintname}
The authors have no competing interests to declare that are
relevant to the content of this article.
\end{credits}
\bibliographystyle{splncs04}
\bibliography{mybibliography}
\end{document}